\pdfoutput=1

\documentclass[twocolumn]{aastex62}
\usepackage{amsmath}
\usepackage{graphicx}
\usepackage{mathrsfs}
\usepackage{color}
\usepackage{url}
\usepackage{CJKutf8}
\usepackage{ulem}

\received{2020}
\revised{\today}
\accepted{2020}

\shorttitle{C/1844 Y1 \& C/2019 Y4}
\shortauthors{Hui \& Ye 2020}

\begin{document}

\title{
Observations of Disintegrating Long-Period Comet C/2019 Y4 (ATLAS) -- A Sibling of C/1844 Y1 (Great Comet)
}

\correspondingauthor{Man-To Hui}
\email{manto@ifa.hawaii.edu}

\author{\begin{CJK}{UTF8}{bsmi}Man-To Hui (許文韜)\end{CJK}}
\affiliation{Institute for Astronomy, University of Hawai`i,
2680 Woodlawn Drive, Honolulu, HI 96822, USA}

\author{\begin{CJK}{UTF8}{bsmi}Quan-Zhi Ye (葉泉志)\end{CJK}}
\affiliation{Department of Astronomy,
University of Maryland,
College Park, MD 20742, USA}


\begin{abstract}

We present a study of C/2019 Y4 (ATLAS) using Sloan $gri$ observations from mid-January to early April 2020. During this timespan, the comet brightened with a growth in the effective cross-section of $\left(2.0 \pm 0.1 \right) \times 10^{2}$ m$^{2}$ s$^{-1}$ from the beginning to $\sim$70 d preperihelion in late March 2020, followed by a brightness fade and the comet gradually losing the central condensation. Meanwhile, the comet became progressively bluer, and was even bluer than the Sun ($g - r \approx 0.2$) when the brightness peaked, likely due to activation of subterranean fresh volatiles exposed to sunlight. With the tailward-bias corrected astrometry we found an enormous radial nongravitational parameter, $A_{1} = \left(+2.25 \pm 0.13\right) \times 10^{-7}$ au d$^{-2}$ in the heliocentric motion of the comet. Taking all of these finds into consideration, we conclude that the comet has disintegrated since mid-March 2020. By no means was the split new to the comet, as we quantified that the comet had undergone another split event around last perihelion $\sim$5 kyr ago, during which its sibling C/1844 Y1 (Great Comet) was produced, with the in-plane component of the separation velocity $\ga$1 m s$^{-1}$. We constrained that the nucleus of C/2019 Y4 before disintegration was $\ga$60 m in radius, and has been protractedly ejecting dust grains of $\sim$10-40 \micron (assuming dust bulk density 0.5 g cm$^{-3}$) with ejection speed $\sim$30 m s$^{-1}$ in early March 2020 and increased to $\sim$80 m s$^{-1}$ towards the end of the month for grains of $\sim$10 \micron.

\end{abstract}


\section{Introduction}

C/2019 Y4 (ATLAS) was a long-period comet that was discovered by the Asteroid Terrestrial-Impact Last Alert System (ATLAS) at Mauna Loa, Hawai`i, on UT 2019 December 28.6.\footnote{See Minor Planet Electronic Circular 2020-A112 (\url{https://minorplanetcenter.net/mpec/K20/K20AB2.html}).} The current orbital solution indicates that the comet orbits around the Sun in a highly elliptical trajectory, with eccentricity $e = 0.999$, perihelion distance $q = 0.25$ au, and inclination $i = 45\fdg4$. Even before the official announcement of the discovery was made by the Minor Planet Center (MPC), amateur astronomer M. Meyer noticed and reported that the orbit of C/2019 Y4 (thence momentarily designated as A10j7UG, when the arc was merely few days) carries a great resemblance to that of C/1844 Y1 (Great Comet), and therefore is a potential sibling of the latter and shares a common progenitor.\footnote{\url{https://groups.io/g/comets-ml/message/28086}}

While as many as $\sim$10$^{5}$ asteroids have been identified to be members of over than a hundred families \citep{2015aste.book..297N}, so far only a small number of comet families have been found, the majority of which consist of only several members except the Kreutz family and the 96P/Machholz complex \citep[e.g.,][]{2004come.book..301B,2005ARA&A..43...75M,2005ApJS..161..551S}. This is probably due to the fact that after disruption, small comets are wiped out as a consequence of their higher susceptibility to rotational instability due to anisotropic mass loss \citep[e.g.,][]{2004come.book..659J,2007AdSpR..39..421S}. The recognition of the genetic relationship between C/1844 Y1 and C/2019 Y4 enriches the comet family samples and is therefore of good value to better understand how comets split and how the family members evolve. 

In this paper, we characterised the physical properties of C/2019 Y4 through Sloan $gri$ observations (Section \ref{sec_obs}), and investigated the genetic relationship between C/1844 Y1 and C/2019 Y4 (Section \ref{sec_split}). The conclusions are summarised in Section \ref{sec_sum}. At the time of writing, our new observations of C/2019 Y4 clearly showed that the nucleus of the comet has split into multiple individual optocentres, and the process is still ongoing. Our detailed analysis of the observed disintegration will be presented in another paper in preparation.

\begin{figure*}
\epsscale{1.0}
\begin{center}
\plotone{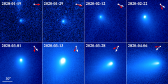}
\caption{
Selected $r$-band coadded images of comet C/2019 Y4 (ATLAS) from NEXT at Xingming Observatory, with intensity stretched in the same logarithmic scale. Dates in UT are labelled. In the lower left a scale bar of 30\arcsec~in length is shown and applicable to all of the panels. The red and white arrows respectively mark the position angles of the antisolar direction and the negative heliocentric velocity vector projected onto the sky plane. Equatorial north is up and east is left. The comet appears slightly trailed in the upper panels because the individual exposure times were longer and the telescope did not track nonsidereally.
\label{fig:19Y4_NEXT}
} 
\end{center} 
\end{figure*}

\begin{figure}
\epsscale{1.15}
\begin{center}
\plotone{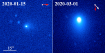}
\caption{
Comet C/2019 Y4 (ATLAS) in $r$-band images in the same logarithmic scale from LDT at Lowell Observatory. Dates in UT are labelled. A scale bar of 15\arcsec~in length is given. The red and white arrows bear the same meanings as in Figure \ref{fig:19Y4_NEXT}. Equatorial north is up and east is left.
\label{fig:19Y4_LDT}
} 
\end{center} 
\end{figure}

\begin{deluxetable*}{lcccccccccc}
\tablecaption{Observing Information and Viewing Geometry of Comet C/2019 Y4 (ATLAS)
\label{tab:vgeo}}
\tablewidth{0pt}
\tablehead{ 
\colhead{Date (UT)} & \colhead{Telescope\tablenotemark{a}} & \colhead{Filter} & \colhead{$t_{\rm exp}$ (s)\tablenotemark{b}} & \colhead{$r_{\rm H}$ (au)\tablenotemark{c}}  & 
\colhead{$\it \Delta$ (au)\tablenotemark{d}} & \colhead{$\alpha$ (\degr)\tablenotemark{e}} & 
\colhead{$\varepsilon$ (\degr)\tablenotemark{f}} &
\colhead{$\theta_{-\odot}$ (\degr)\tablenotemark{g}} &
\colhead{$\theta_{-{\bf V}}$ (\degr)\tablenotemark{h}} &
\colhead{$\psi$ (\degr)\tablenotemark{i}}
}
\startdata
2020 Jan 15\tablenotemark{$\dagger$} & LDT & {\it g}, {\it r}, {\it i} & 180 & 2.685 & 1.998 & 17.5 & 124.9 & 272.4 & 268.7 & 2.1\\
2020 Jan 19 & NEXT & {\it g}, {\it r}, {\it i} & 210 & 2.623 & 1.898 & 17.2 & 128.2 & 267.7 & 266.6 & 0.6 \\
2020 Jan 29 & NEXT & {\it g}, {\it r}, {\it i} & 180 & 2.479 & 1.688 & 16.5 & 134.3 & 253.5 & 260.5 & -3.5 \\
2020 Feb 02 & NEXT & {\it g}, {\it r}, {\it i} & 150 & 2.418 & 1.609 & 16.4 & 136.1 & 246.1 & 257.3 & -5.5 \\
2020 Feb 12 & NEXT & {\it r} & 150 & 2.270 & 1.441 & 17.0 & 137.6 & 225.1 & 247.5 & -10.9 \\
2020 Feb 20 & NEXT & {\it g}, {\it r}, {\it i} & 150 & 2.147 & 1.328 & 18.9 & 135.3 & 206.1 & 237.0 & -15.8 \\
2020 Feb 21 & NEXT & {\it g}, {\it r}, {\it i} & 150 & 2.132 & 1.316 & 19.2 & 134.8 & 203.8 & 235.6 & -16.4 \\
2020 Feb 22 & NEXT & {\it g}, {\it r}, {\it i} & 150 & 2.116 & 1.304 & 19.6 & 134.3 & 201.5 & 234.1 & -17.0 \\
2020 Mar 01 & LDT & {\it r} & 30 & 1.994 & 1.220 & 22.9 & 128.5 & 183.3 & 220.8 & -22.1 \\
2020 Mar 01 & NEXT & {\it g}, {\it r}, {\it i} & 90 & 1.989 & 1.217 & 23.1 & 128.2 & 182.6 & 220.2 & -22.3 \\
2020 Mar 13 & NEXT & {\it g}, {\it r}, {\it i} & 90 & 1.795 & 1.126 & 30.0 & 115.5 & 154.6 & 194.2 & -30.0 \\
2020 Mar 27 & NEXT & {\it g}, {\it r}, {\it i} & 30 & 1.550 & 1.060 & 39.6 & 97.7 & 118.9 & 156.7 & -38.0 \\
2020 Mar 28 & NEXT & {\it g}, {\it r}, {\it i} & 30 & 1.533 & 1.056 & 40.3 & 96.5 & 116.5 & 154.2 & -38.5 \\
2020 Mar 29 & NEXT & {\it g}, {\it r}, {\it i} & 30 & 1.515 & 1.053 & 41.0 & 95.2 & 114.0 & 151.5 & -39.0 \\
2020 Mar 30 & NEXT & {\it g}, {\it r}, {\it i} & 30 & 1.496 & 1.049 & 41.8 & 93.8 & 111.3 & 148.6 & -39.5 \\
2020 Mar 31 & NEXT & {\it g}, {\it r}, {\it i} & 30 & 1.478 & 1.046 & 42.5 & 92.5 & 109.0 & 146.1 & -39.9 \\
2020 Apr 01 & NEXT & {\it g}, {\it r}, {\it i} & 30 & 1.459 & 1.042 & 43.2 & 91.2 & 106.5 & 143.4 & -40.4 \\
2020 Apr 02 & NEXT & {\it g}, {\it r}, {\it i} & 30 & 1.440 & 1.039 & 44.0 & 89.9 & 104.0 & 140.8 & -40.9 \\
2020 Apr 05 & NEXT & {\it g}, {\it r}, {\it i} & 30 & 1.389 & 1.030 & 46.0 & 86.3 & 97.7 & 134.0 & -42.0 \\
2020 Apr 06 & NEXT & {\it g}, {\it r}, {\it i} & 30 & 1.370 & 1.027 & 46.7 & 85.0 & 95.5 & 131.6 & -42.4
\enddata
\tablenotetext{a}{NEXT = 0.6 m Ningbo Education Xinjiang Telescope, LDT = 4.3 m Lowell Discovery Telescope.}
\tablenotetext{b}{Individual exposure time.}
\tablenotetext{c}{Heliocentric distance.}
\tablenotetext{d}{Topocentric distance.}
\tablenotetext{e}{Phase angle (Sun-comet-observer).}
\tablenotetext{f}{Solar elongation (Sun-observer-comet).}
\tablenotetext{g}{Position angle of projected antisolar direction.}
\tablenotetext{h}{Position angle of projected negative heliocentric velocity of the comet.}
\tablenotetext{i}{Observer to comet's orbital plane angle with vertex at the comet. Negative values indicate observer below the orbital plane of the comet.}
\tablenotetext{\dagger}{Sloan $z$ images of the comet were obtained in addition to the $g$, $r$ and $i$-band images on 2020 January 15. However, since there was no additional $z$ images from other epochs, we omit this band throughout this paper.}
\end{deluxetable*}

\begin{figure*}
\gridline{\fig{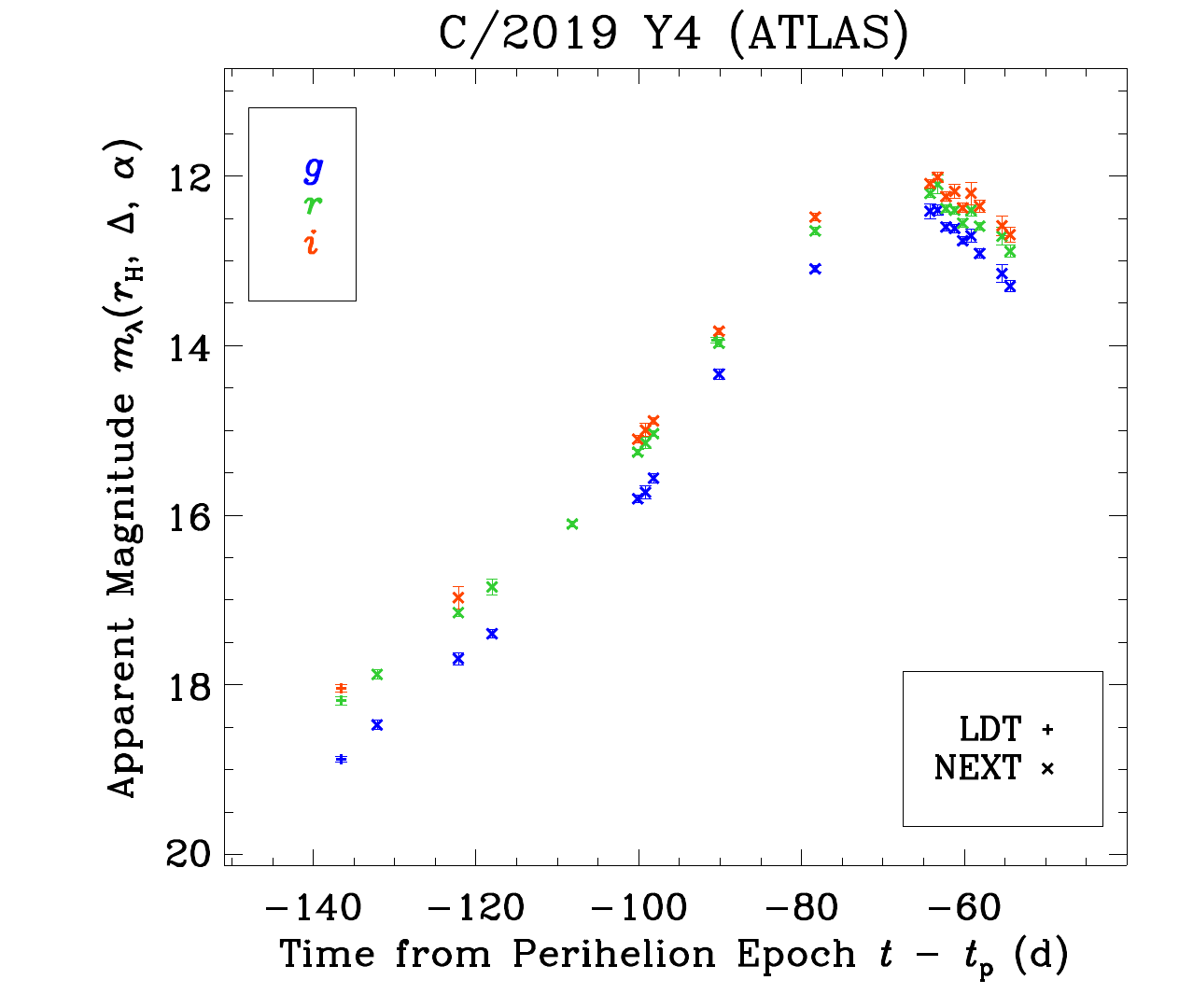}{0.5\textwidth}{(a)}
          \fig{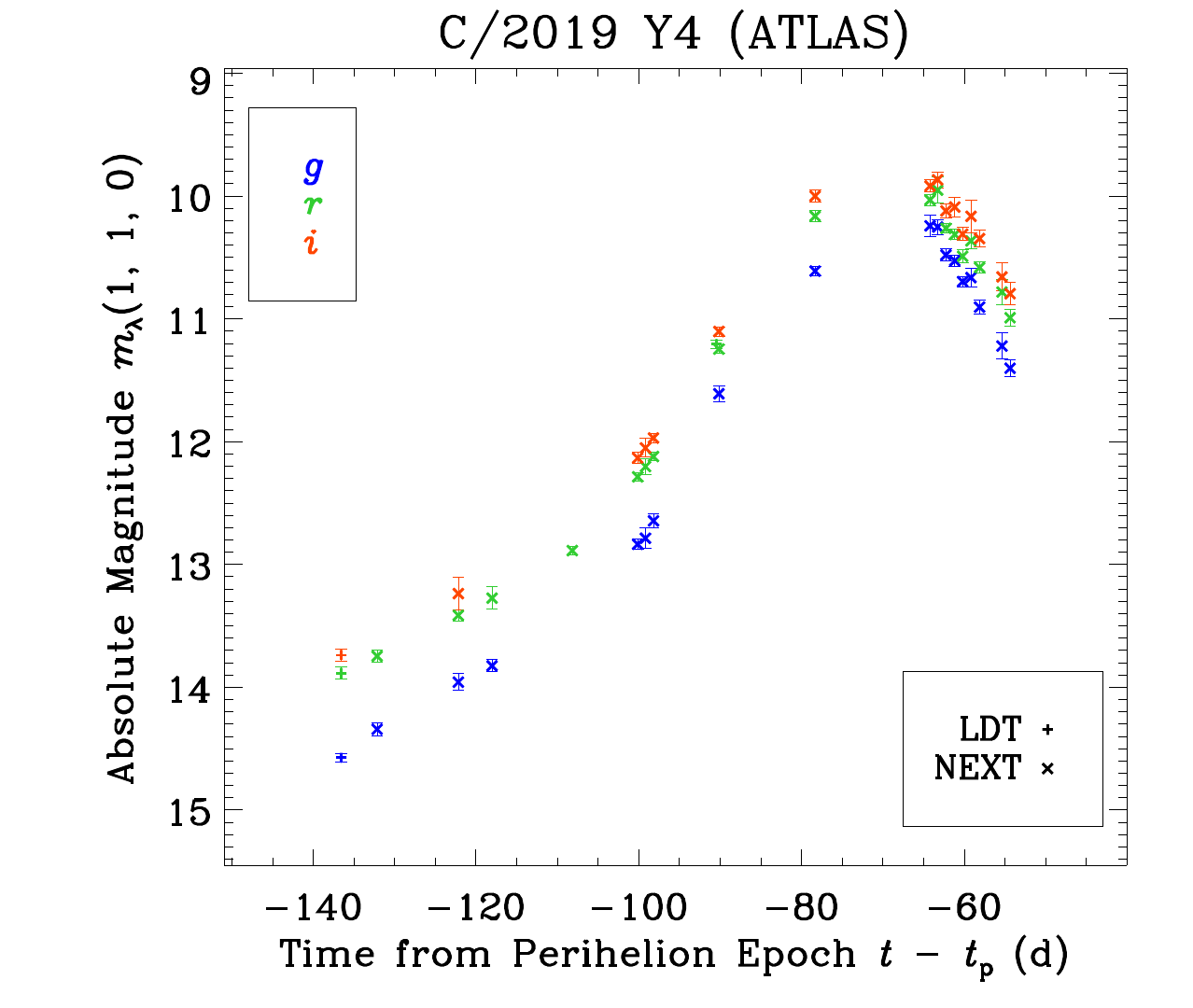}{0.5\textwidth}{(b)}
          }
\caption{
The apparent (a) and intrinsic (b) Sloan $g$, $r$ and $i$-band lightcurves of comet C/2019 Y4 (ATLAS) as functions of time (in terms of time from the perihelion epoch of the comet, $t_{\rm p} = $ TDB 2020 May 31.0) from NEXT (cross) and LDT (plus). Datapoints from different filters are colour coded as indicated in the legend. Equation (\ref{eq_m_abs}) was applied to obtain panel (b) from panel (a). See Section \ref{ssec_lc} for details. During the observed timespan, the comet brightened intrinsically until $\sim$70 days prior to the perihelion, whereafter a decline in brightness was seen.
\label{fig:lc}
} 
\end{figure*}

\section{Observation}
\label{sec_obs}


We use the publicly available images of comet C/2019 Y4 taken by the Ningbo Education Xinjiang Telescope (NEXT), which is a 0.6 m telescope located at Xingming Observatory, Xinjiang, China. Regular monitoring of C/2019 Y4 started on 2020 January 19 and continued to the start of this project. Images were taken with a 2k$\times$2k CCD mostly through Sloan $g$, $r$, and $i$ filters, yet in a few nights only $r$-band images were obtained. As the telescope did not follow the nonsidereal motion, a slight trailing of the comet can be noticed in images from January to February 2020 (see Figure \ref{fig:19Y4_NEXT}), when longer individual exposures were used. The images have an unbinned pixel scale of 0\farcs63, with a field-of-view of $22'\times22'$, and a typical full-wide-half-maximum (FWHM) of 2\arcsec-3\arcsec. We employed {\tt AstroImageJ} \citep{2017AJ....153...77C} to subtract bias and dark frames from the images, which were subsequently divided by flat frames. Then we derived plate constants of the images with the Gaia DR2 catalog \citep{2018A&A...616A...1G}; the photometric image zeropoints were derived with the Pan-STARRS DR1 catalog \citep{2013ApJS..205...20M} using field stars with sun-like colors (defined as color indices within $\pm$0.2 mag from the solar value). The zeropoints were then converted to the SDSS photometric system using the relation derived in \citep{2012ApJ...750...99T}. The procedure was performed with {\tt PHOTOMETRYPIPELINE} \citep{2017A&C....18...47M}.


Additional Sloan $g$, $r$, and $i$-band images of C/2019 Y4 were obtained with the Large Monolithic Imager \citep[LMI;][]{2013AAS...22134502M} on the 4.3 m Lowell Discovery Telescope (LDT; formerly known as the Discovery Channel Telescope) tracking nonsidereally on 2020 January 15 and March 01. These images have a field-of-view of $12\farcm3 \times 12\farcm3$, with a pixel scale of 0\farcs24 after a $2\times2$ on-chip binning, and a typical FWHM of $\sim$1\arcsec~for the field stars. We handled and photometrically calibrated the LDT images following exactly the same procedures that we applied to the NEXT images. Figure \ref{fig:19Y4_LDT} shows two of the individual $r$-band images of the comet from the two nights at LDT.

We summarise the observations and the viewing geometry of C/2019 Y4 from NEXT and LDT in Table \ref{tab:vgeo}.

\subsection{Lightcurve \& Colour}
\label{ssec_lc}

We took measurements of comet C/2019 Y4 in the NEXT and LDT images using an aperture of fixed linear radius $\varrho = 10^{4}$ km projected at the distance of the comet from the optocentre. The equivalent apparent angular size of the aperture is always large enough such that the slight trailing of the comet in the NEXT data from January to February 2020 would not be a concern. Figure \ref{fig:lc}a shows our multiband lightcurve measurements as functions of time, in terms of time from the epoch of perihelion passage of C/2019 Y4 ($t_{\rm p} = $ TDB 2020 May 31.0). The comet apparently brightened on its way to perihelion in a continuous manner until $t - t_{\rm p} \ga -70$ d, after which the downtrend in brightness was seen.

\begin{figure*}
\gridline{\fig{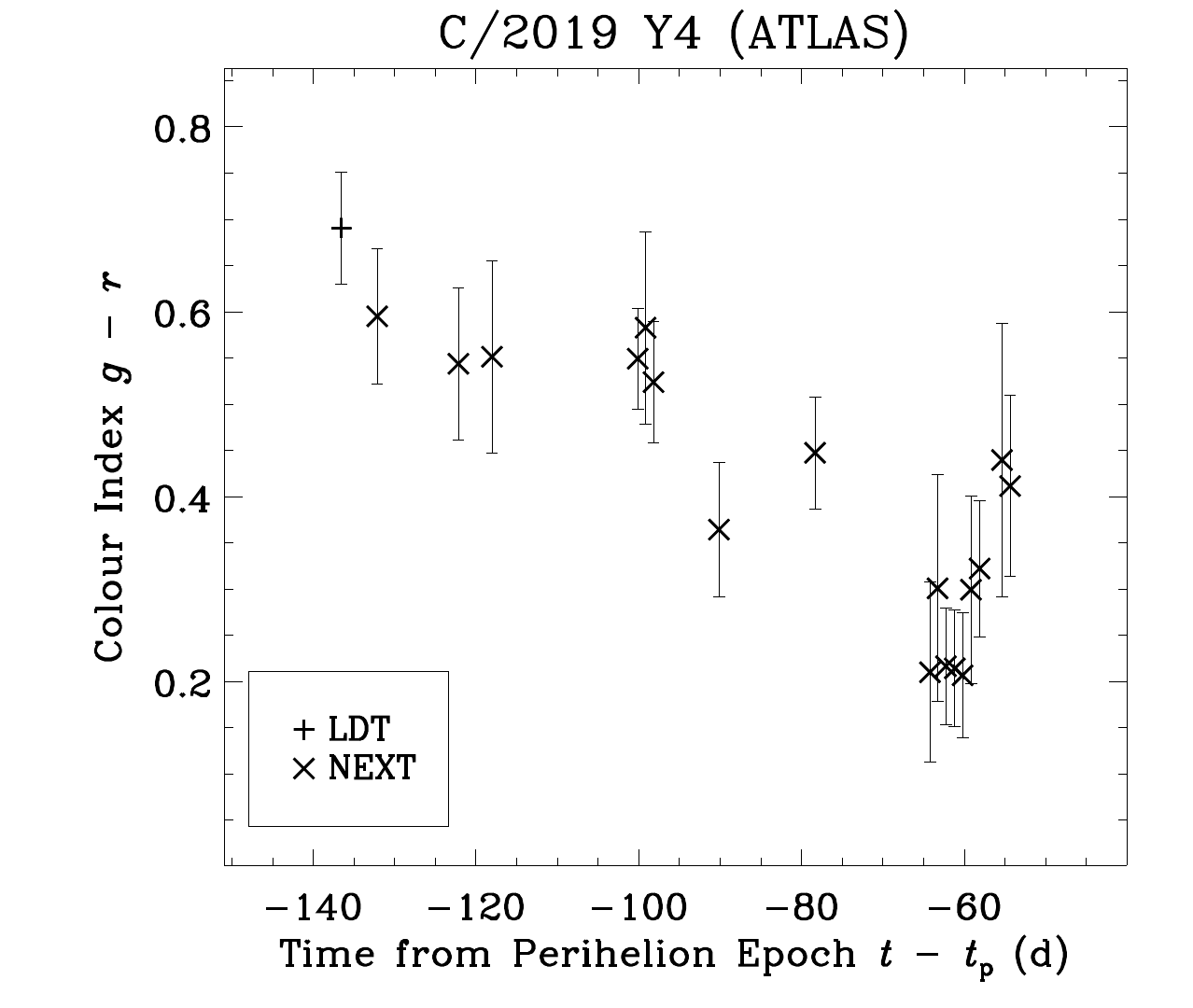}{0.5\textwidth}{(a)}
          \fig{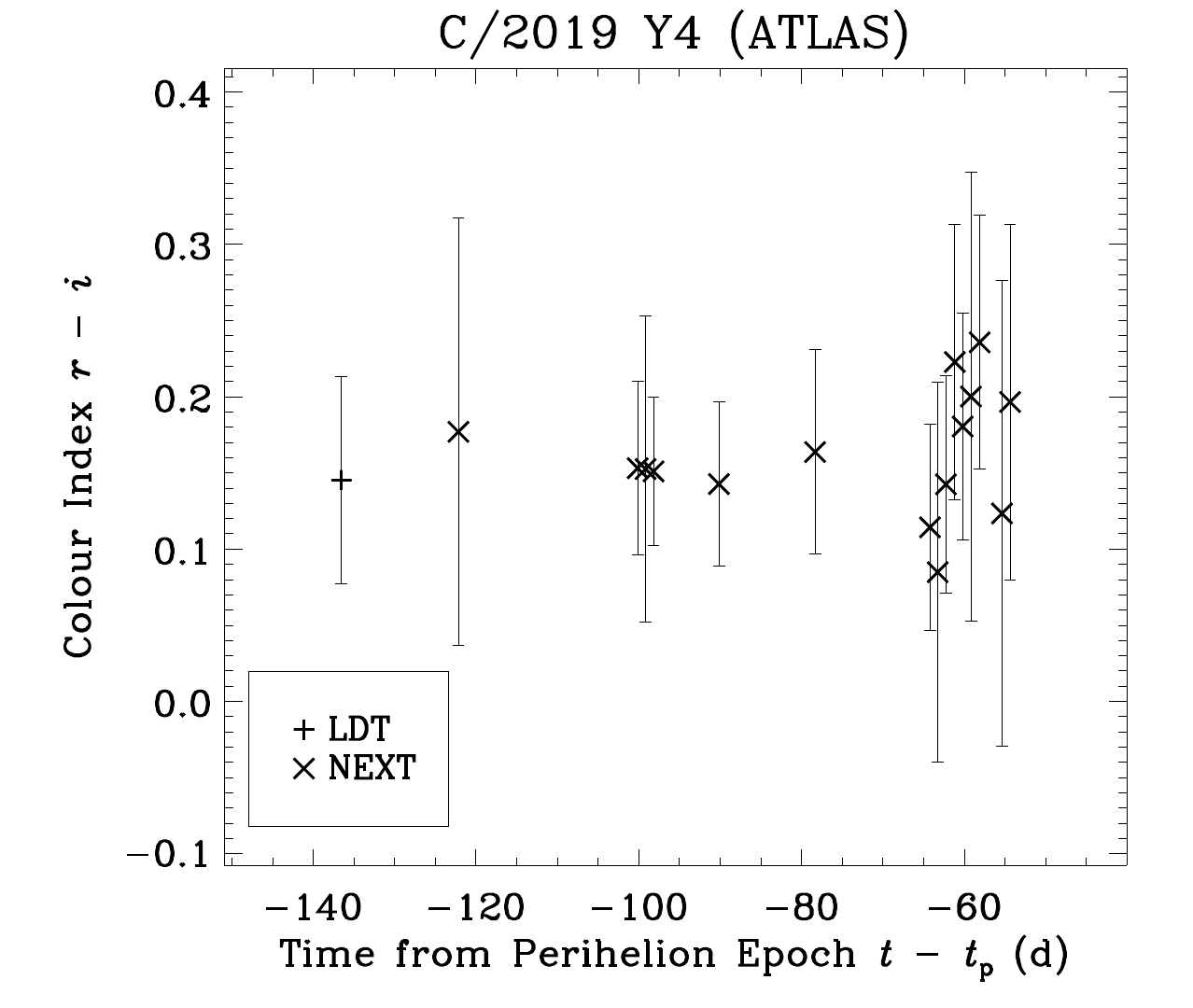}{0.5\textwidth}{(b)}
          }
\caption{
The temporal evolution of the colour of comet C/2019 Y4 (ATLAS) in terms of colour indices (a) $g - r$ and (b) $r-i$ from NEXT (cross) and LDT (plus). During the observed timespan, while no statistically confident $r - i$ variation was witnessed, we can notice a blueing trend in the $g - r$ colour until $\sim$60 d preperihelion, after which the comet appeared to be reddening.
\label{fig:clr}
} 
\end{figure*}

We also show the colour of the comet in terms of $g - r$ and $r - i$ colour indices respectively in the left and right panels of Figure \ref{fig:clr}. Interestingly, while the $r - i$ colour index of the comet remained constant and sun-like \citep[i.e., $\left(r - i\right)_{\odot} = +0.12 \pm 0.02$;][]{2018ApJS..236...47W} given the measurement uncertainties, the colour across the $g$ and $r$ bands seems to indicate that the comet had a blueing trend from a reddish colour \citep[$g - r \approx 0.6$, in comparison to the Sun's $\left(g - r\right)_{\odot} = +0.46 \pm 0.04$;][]{2018ApJS..236...47W} since January 2020, reached a dip at an epoch of $\sim$60 d preperihelion, when the comet appeared even bluer than the Sun ($g - r \approx 0.2$), and began to be reddening afterwards. Accordingly, we argue that the blueing dip was caused by gas emission from a massive amount of previously buried fresh volatiles suddenly exposed to the sunlight, indicative of a disintegration event in mid-March 2020.

We evaluated the intrinsic lightcurve of the comet by correcting the varying observing geometry and computed its absolute magnitude from the apparent magnitude from
\begin{equation}
m_{\lambda} \left(1, 1, 0 \right) = m_{\lambda} \left(r_{\rm H}, {\it \Delta}, \alpha \right) - 5 \log \left(r_{\rm H} {\it \Delta} \right) + 2.5 \log \phi \left(\alpha\right)
\label{eq_m_abs},
\end{equation}
\noindent in which $\lambda$ is the magnitude bandpass, $r_{\rm H}$ and ${\it \Delta}$ are the heliocentric and topocentric distances, respectively, both expressed in au, and $\phi \left(\alpha\right)$ is the phase function of the comet, approximated by the empirical Halley-Marcus phase function \citep{2007ICQ....29...39M,2011AJ....141..177S}. The resulting intrinsic lightcurve of C/2019 Y4 is plotted in Figure \ref{fig:lc}b, from which we clearly notice that the lightcurve trend appears broadly the same as the one in Figure \ref{fig:lc}a. Indeed the comet continuously brightened until $\la$70 d preperihelion, thereafter followed by a conspicuous fading process in an intrinsic manner, which we think also acts as a piece of evidence that a disintegration event has occurred to comet C/2019 Y4.

\subsection{Activity \& Nucleus Size}
\label{ssec_act}

The change in the intrinsic brightness is mostly related to the variation in the effective scattering geometric cross-section of the dust particles through
\begin{equation}
C_{\rm e} = \frac{\pi r_{0}^{2}}{p_{r}} 10^{0.4 \left[m_{\odot,r} - m_{r} \left(1,1,0 \right) \right]}
\label{eq_XS},
\end{equation}
\noindent where $C_{\rm e}$ is the cross-section, $m_{\odot, r} = -26.93$ is the apparent {\it r}-band magnitude of the Sun at the mean Earth-Sun distance $r_{0} = 1.5 \times 10^{8}$ km \citep{2018ApJS..236...47W}, and $p_{r} = 0.1$ is the assumed value for the $r$-band geometric albedo of cometary dust \citep{2017JQSRT.202..104Z}, as the true value remains unconstrained. The reason why we only focus on the $r$-band data here is that these images have less contamination from gaseous emission than the $g$-band ones do, and they have higher sensitivity than the $i$-band images do. By employing Equation (\ref{eq_XS}), we estimated the change in the effective cross-section within the fixed photometric aperture during the brightening part ($-140 \la t - t_{\rm p} \la -80$ d) of the lightcurve to be $\Delta C_{\rm e} = \left(9.9 \pm 0.5 \right) \times 10^{2}$ km$^{2}$, corresponding to an average growth rate in the effective cross-section of $\left\langle \dot{C}_{\rm e} \right\rangle = \left(2.0 \pm 0.1 \right) \times 10^{2}$ m$^{2}$ s$^{-1}$. Assuming that the increased cross-section consists of dust grains with mean radius $\bar{\mathfrak{a}}$ and bulk density $\rho_{\rm d}$, the average net mass-loss rate within the photometric aperture is then given by 
\begin{equation}
\left\langle \dot{M}_{\rm d}\right\rangle = \frac{4}{3} \rho_{\rm d} \bar{\mathfrak{a}} \left\langle \dot{C}_{\rm e}\right\rangle
\label{eq_mloss}.
\end{equation}
\noindent The product of $\bar{\mathfrak{a}}$ and $\rho_{\rm d}$ is inversely proportional to the $\beta$ parameter ($0.03 \la \beta \la 0.1$, see Section \ref{ssec_morph}). Substitution into Equation (\ref{eq_mloss}) gives us $\left\langle \dot{M}_{\rm d}\right\rangle \approx 4 \pm 2$ kg s$^{-1}$ for C/2019 Y4 during the observed brightening period.

An approach to constrain the nucleus size of C/2019 Y4 is to estimate the minimum active surface that would be needed to supply the mass-loss rate during the brightening process, provided that the activity was all driven by sublimation of water (H$_{2}$O) ice. The corresponding lower bound to the nucleus size can then be estimated from
\begin{equation}
R_{\rm n} \ga \sqrt{\frac{\left\langle \dot{M}_{\rm d} \right \rangle}{\pi f_{\rm s}}}
\label{eq_Rnuc2}.
\end{equation}
\noindent The equilibrium sublimation mass flux of H$_{2}$O gas would be $1.2 \times 10^{-6} \la f_{\rm s} \la 1.2 \times 10^{-4}$ kg s$^{-1}$ m$^{-2}$ at the range of the heliocentric distances during the timespan ($1.8 \la r_{\rm H} \la 2.7$ au). Inserting numbers in, we obtain $R_{\rm n} \ga 60$ m for the nucleus size of the comet.

An upper limit to the nucleus size of C/2019 Y4 could have been derived from our detection of the nongravitational acceleration (Section \ref{ssec_od}). However, strictly speaking, the nongravitational effect is only applicable to the barycentre of the unresolved fragments, rather than an intact nucleus. The equivalent bulk density of the barycentre should be much lower than those of typical cometary nuclei \citep[e.g., $533 \pm 6$ kg m$^{-3}$ for 67P/Churyumov–Gerasimenko;][]{2016Natur.530...63P}, to a degree we cannot firmly constrain. We thus posit that applying this approach by assuming a typical bulk density for cometary nuclei is no longer valid.

\begin{figure*}
\epsscale{1.0}
\begin{center}
\plotone{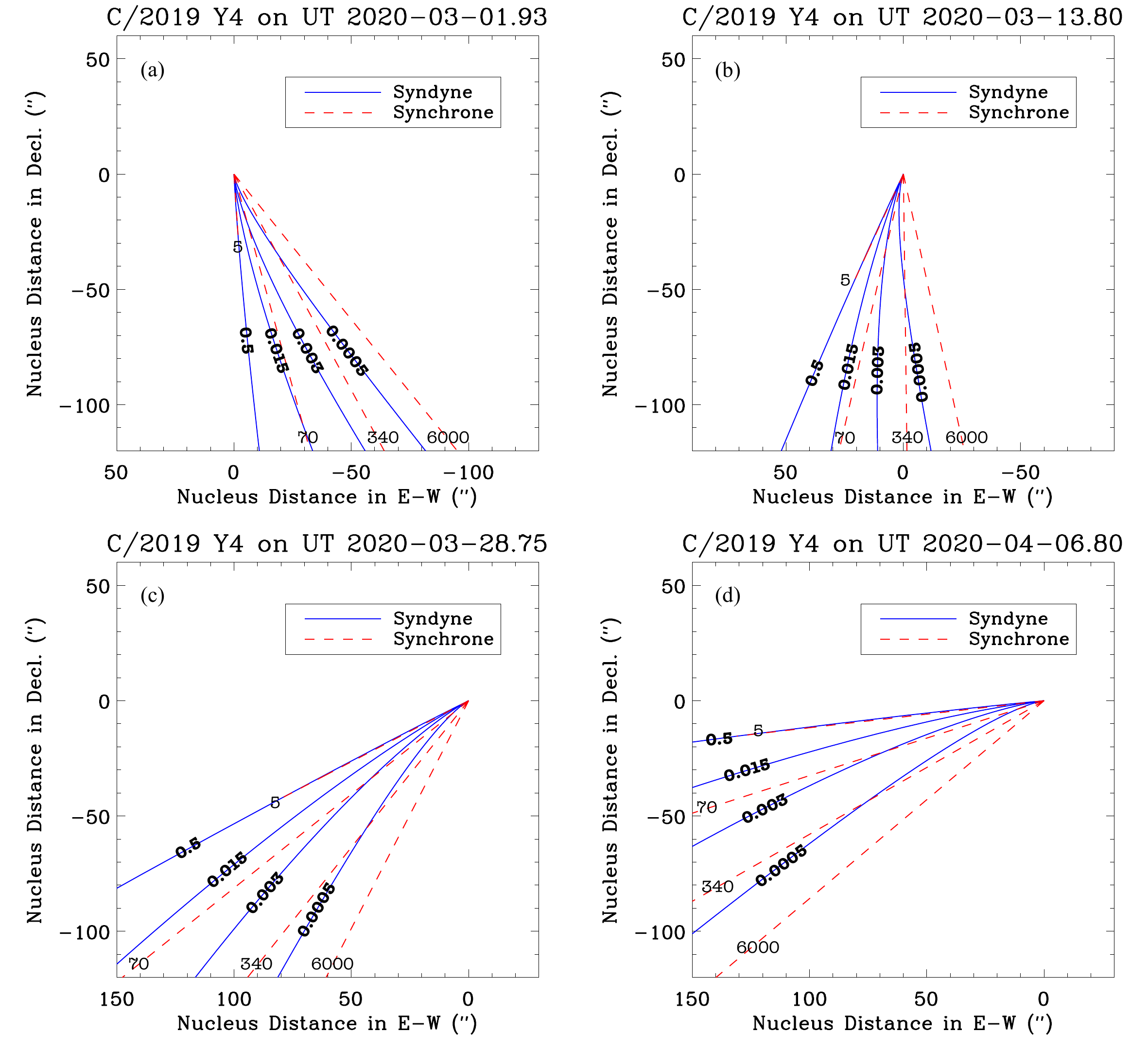}
\caption{
Examples of syndyne-synchrone grids for comet C/2019 Y4 (ATLAS) on (a) UT 2020 March 01, (b) March 13, (c) March 28 and (d) April 6. As pointed out by the legend in each panel, the syndynes are plotted as blue curves, with the values of the $\beta$ parameter labelled as bold texts, and the synchrones are plotted as red dashed curves, with the grain release time from the observed epochs and expressed in days labelled as horizontally oriented unbolded texts.
\label{fig:19Y4_FP}
} 
\end{center} 
\end{figure*}

\subsection{Morphology}
\label{ssec_morph}

C/2019 Y4 has been unambiguously cometary since our earliest observation in January 2020. Its central optocentre had been strongly condensed until 2020 April 05, after which it started to become even more diffuse and elongated (see Figure \ref{fig:19Y4_NEXT}).\footnote{The coadded image from 2020 April 05 is not shown, yet the appearance of the comet is similar to that on April 06.} The morphological change, altogether with ongoing observations from NEXT in which we see multiple optocentres in the coma, strongly indicates that the cometary nucleus has split into multiple pieces of fragments \citep{2020ATel13620....1Y}. This feature has been confirmed by \citet{2020ATel13622....1S} and \citet{2020ATel13629....1S}. A detailed study about the disintegration and followup observations will be presented in another paper in preparation. 

Physical properties of cometary dust can be revealed by studying the morphology of cometary dust tails\citep[e.g.,][]{2004come.book..565F}. To better understand comet C/2019 Y4, here we adopted the classical syndyne-synchrone computation \citep[e.g.][]{1968ApJ...154..327F}. A syndyne line is loci of dust grains that are subject to the $\beta$ parameter in common, which is the ratio between the solar radiation pressure force and the gravitational force of the Sun, and is inversely proportional to the product of the dust bulk density and the grain radius, $\rho_{\rm d} \mathfrak{a}$, but are freed from the nucleus at various release epochs. Dust grains that are driven by different values of $\beta$ and are released from the nucleus at a common epoch constitute a synchrone line. In the syndyne-synchrone approximation, the ejection velocity of dust grains is ignored.

We focused on the $r$-band images taken since 2020 March from NEXT, in which the dust tail of C/2019 Y4 was recorded the most clearly and the optocentre appeared untrailed. In Figure \ref{fig:19Y4_FP}, we plot four examples of the syndyne-synchrone grid computation. Here for better visualisation, only sparse syndyne-synchrone grids are presented. A much denser grid was used to visually determine the $\beta$ parameter and the dust release epochs about which the dust tail appeared to be symmetrical. For data from some single observed epoch, we find it difficult to judge whether the symmetry of the dust tail appeared more like a syndyne or synchrone. However, with results from multiple observed epochs, we realised that the dust tail of the comet can be better approximated by syndyne lines with $0.03 \la \beta \la 0.1$, because the range of the $\beta$ parameter remained roughly the same, whereas the dust release epoch would keep changing, were the tail closer to a synchrone line. This find is consistent with the fact that the comet has been active protractedly since the discovery. Assuming a typical value of $\rho_{\rm d} = 0.5$ g cm$^{-3}$ for the bulk density of cometary dust of C/2019 Y4, we find the grain radius to be $10 \la \bar{\mathfrak{a}} \la 40$ \micron, fully within the known dust-size range of other long-period comets \citep{2004come.book..565F}.

As the syndyne-synchrone computation does not unveil the ejection speed of the observed dust grains of the comet, we estimate this quantity, denoted as $V_{\rm ej}$, by means of measuring the apparent length of the sunward extent to the dust coma $\ell$. The two quantities are connected by the following relationship
\begin{equation}
V_{\rm ej} = \frac{\sqrt{2 \beta \mu_{\odot} {\it \Delta} \tan \ell \sin \alpha}}{r_{\rm H}}
\label{eq_vej},
\end{equation}
\noindent where $\mu_{\odot} = 3.96 \times 10^{-14}$ au$^{3}$ s$^{-2}$ is the heliocentric gravitational constant. We found $\ell \approx 10\arcsec$ on 2020 March 01, $\sim$15\arcsec~on March 13, and $\sim$25\arcsec~on March 31. Substituting, Equation (\ref{eq_vej}) yields $V_{\rm ej} \approx 30$ m s$^{-1}$ at the beginning of March 2020, $\sim$50 m s$^{-1}$ around halfway, and further increased to $\sim$80 m s$^{-1}$ at the end of the month for dust grains of $\beta \sim 0.1$. This find is similar to what has been identified for other long-period comets \citep[e.g.,][]{2014ApJ...791..118M}.

\section{Fragmentation of the Parent}
\label{sec_split}

\subsection{Orbit Determination}
\label{ssec_od}

\begin{figure*}
\epsscale{1.0}
\begin{center}
\plotone{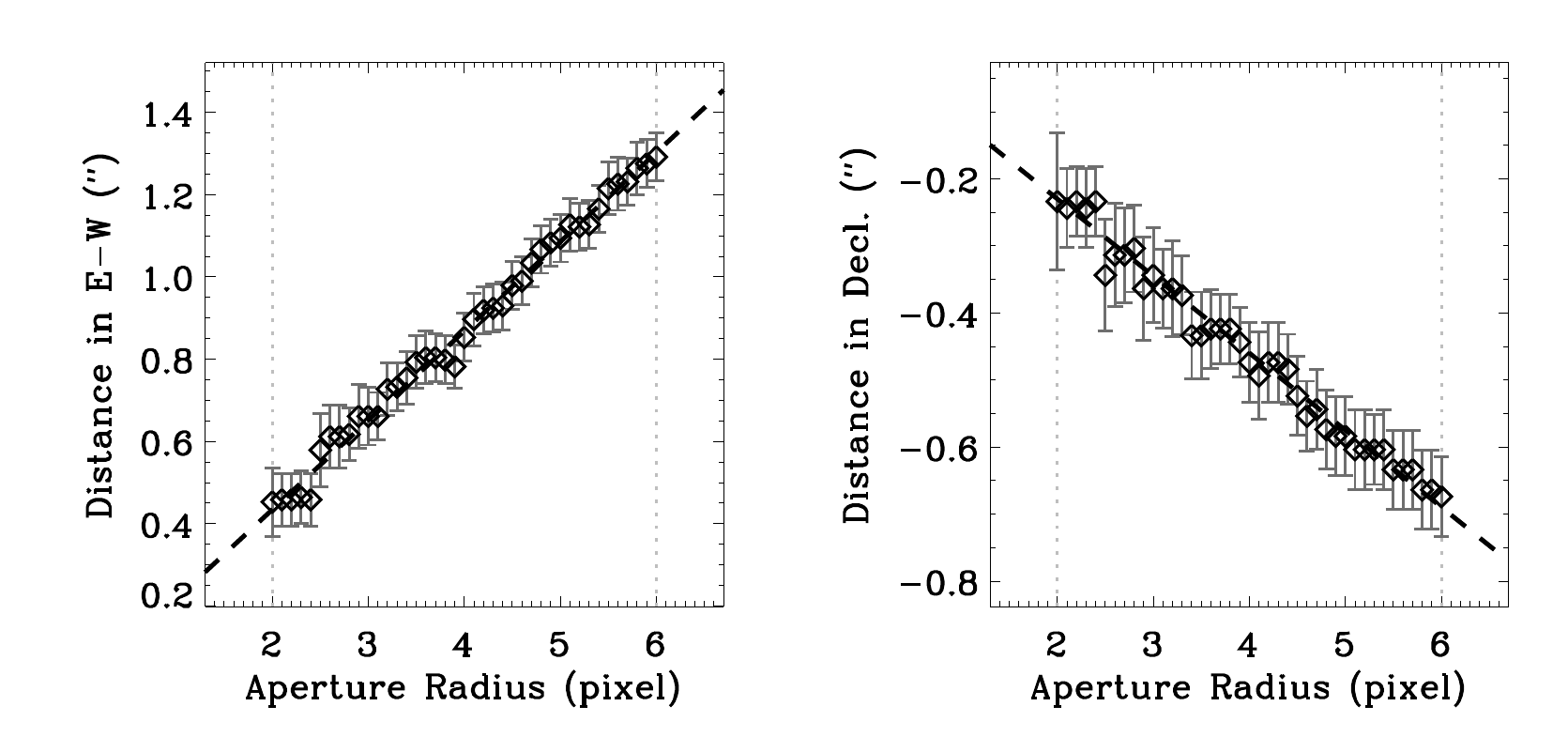}
\caption{
Example of the angular distance from the zero-aperture centroid in the J2000 equatorial east-west (left) and declination (right) directions as functions of the astrometric aperture size for comet C/2019 Y4 (ATLAS) in a NEXT {\it r}-band image from 2020 March 29. The dashed lines are the best-fit least-square linear functions to the datapoints. In each panel, the two vertical dotted lines mark the range of aperture radii used for the best fits.
\label{fig:19Y4_twb}
} 
\end{center} 
\end{figure*}

We performed astrometry of comet C/2019 Y4 in the {\it r}-band images from LDT and NEXT by exploiting AstroMagic\footnote{\url{http://www.astromagic.it/eng/astromagic.html}} and codes developed by D. Tholen with the Gaia DR2 catalogue \citep{2018A&A...616A...1G}. In this step, we realised that the tailward bias of the astrometric measurements was readily conspicuous in most of the observed images before early April 2020, after which the comet apparently lost the central condensation visibly and so no astrometry was measured.\footnote{Actually we have measured a number of images from 2020 April 05 and 06 as a test. However, the measurement uncertainties reach $\ga$1\arcsec, and therefore we decided not to include these measurements or to continue measuring astrometry of C/2019 Y4.} To make sure that our orbit determination will not be skewed by the tailward bias in the astrometry, we performed least-square linear fits to the centroids in the J2000 equatorial east-west and declination directions, respectively, as functions of the aperture size (Figure \ref{fig:19Y4_twb}). The zero-aperture astrometry was then obtained, with its uncertainties propagated from the centroiding errors. In addition to our astrometric measurements of C/2019 Y4, we also included the astrometric observations of the comet from station T12 (Tholen NEO Follow-Up at the University of Hawai`i 2.24 m telescope) available from the MPC Observations Database\footnote{\url{https://minorplanetcenter.net/db_search}}, which have been corrected for the tailward bias as well (D. Tholen, from whom we obtained the corresponding measurement errors through private communication). The other astrometric observations available from the MPC Observations Database had to be discarded, because we do not think that they are zero-aperture astrometry of the comet, and no astrometric uncertainty is available either.

We employed the orbit determination code {\tt FindOrb}\footnote{\url{https://www.projectpluto.com/find_orb.htm}} developed by B. Gray, which incorporates gravitational perturbation from the eight major planets, Pluto, the Moon, and the most massive 16 main-belt asteroids. The code applies post-Newtonian corrections, and uses the planetary and lunar ephemerides DE431 \citep{2014IPNPR.196C...1F}. Initially we attempted to fit a purely gravitational orbit to the astrometric observations, however, the resulting astrometric residuals exhibits an obvious systematic trend in observations beyond $3\sigma$ from both the beginning and the end of the observed arc. The mean RMS residual of the fit is 0\farcs266 from 104 observations in total. However, after we included the radial and transverse nongravitational parameters $A_{1}$ and $A_{2}$, first introduced by \citet{1973AJ.....78..211M}, in a nongravitational force model in which the nongravitational acceleration of the nucleus is assumed to be proportional to the mass flux of hemispherical H$_{2}$O-ice sublimation (Hui \& Farnocchia, in preparation), the trend can be completely removed and the mean RMS residual of the fit shrinks to 0\farcs126, only roughly the half of the one in the gravity-only solution. Adding the normal component of the nongravitational parameter $A_{3}$ does not improve the orbital fit. Nor is the obtained $A_{3}$ statistically significant. Furthermore, only in very few cases has $A_{3}$ been determined meaningfully \citep{2004come.book..137Y}, suggesting that $A_{3}$ plays a less significant role in comparison to $A_{1}$ and $A_{2}$. Therefore, we opted not to include $A_{3}$ but only $A_{1}$ and $A_{2}$. Our best-fit nongravitational solution to the orbit of C/2019 Y4 as well as the associated details are summarised in Table \ref{tab:orb}, where we can see that the radial nongravitational parameter of the comet, $A_{1} = \left(+2.25 \pm 0.13 \right) \times 10^{-7}$ au d$^{-2}$, is particularly enormous amongst the whole comet population, but is by no means unseen amongst disintegrated comets, e.g., $A_{1} = \left(+1.21 \pm 0.12\right) \times 10^{-6}$ au d$^{-2}$ for C/2015 D1 (SOHO) by \citet{2015ApJ...813...73H}, and $A_{1} = \left(+1.74 \pm 0.11 \right) \times 10^{-7}$ au d$^{-2}$ for C/2017 E4 (Lovejoy) by JPL Horizons. The obtained transverse nongravitational parameter, $A_{2} = \left(-3.1 \pm 1.0\right) \times 10^{-8}$ au d$^{-2}$, is far less significant than its radial counterpart $A_{1}$ by almost an order of magnitude, and yet is nevertheless typical in the context of disintegrated comets, e.g., $A_{2} = \left(-1.55 \pm 0.09 \right) \times 10^{-8}$ au d$^{-2}$ for C/1999 S4 (LINEAR), and $A_{2} = \left(+6.2 \pm 0.8 \right) \times 10^{-8}$ au d$^{-2}$ for C/2010 X1 (Elenin), both computed by JPL Horizons. 

Our result is similar to the nongravitational solutions to the orbit of C/2019 Y4 by the MPC\footnote{\url{https://minorplanetcenter.net/mpec/K20/K20H28.html}} ($A_{1} = +2.6 \times 10^{-7}$ au d$^{-1}$ and $A_{2} = -2.9 \times 10^{-8}$ au d$^{-1}$, no uncertainties given) and by JPL Horizons ($A_{1} = \left(+2.86 \pm 0.17 \right) \times 10^{-7}$ au d$^{-2}$ and $A_{2} = \left(-0.9 \pm 1.2 \right) \times 10^{-8}$ au d$^{-2}$), despite that most of their used astrometric observations are likely uncorrected for the tailward bias, and the old nongravitational force model by \citet{1973AJ.....78..211M} was adopted. We think that in this specific case, although the tailward bias is obviously present in their used data from individual nights, it has been fortuitously negated by the varying observing geometry (Table \ref{tab:vgeo}), as well as by the enormous nongravitational effect.

In order to investigate the dynamical relationship between C/1844 Y1 and C/2019 Y4, we also derived a gravity-only orbit for the latter, using a shorter observed arc during which we found no statistically significant nongravitational effect, because we prefer that the enormity of the nongravitational effect is unlikely to be characteristic of the complete orbit, but reflects the ongoing disintegration in the current apparition. For this purpose, we only wanted to include datapoints from the period when the nongravitational force has not yet played an important role. The earliest astrometric data from T12 were always included. We tested with both gravity-only and nongravitational force models, and checked astrometric residuals of the astrometry, the significance of the radial nongravitational parameter $A_{1}$, and the mean RMS residuals of the fits. What we found is that, if any astrometric observations from 2020 March 13 and thereafter are included, the nongravitational solution to the orbit of C/2019 Y4 improves the fit considerably in comparison to the gravity-only version. Thus, by discarding all of the astrometric observations starting from 2020 March 13, we obtained the final version of the gravity-only solution to the orbit of the comet. The information is summarised in Table \ref{tab:orb} as well.

For C/1844 Y1, neither the MPC nor JPL Horizons give the uncertainty information of its orbital elements. Thus, we extracted the topocentric astrometry from \citet{1850AJ......1...97B}. Only observations with both R.A. and decl. measurements available from the same epochs were used and fed into {\tt FindOrb}. The equatorial coordinates were precessed from epoch 1845.0 to 2000.0. Based upon the mean residual of the preliminary orbital solution, we downweighted all of the observations by an equal uncertainty of 15\arcsec. Ten (out of 80 in total) observations with astrometric residuals $\ga 3\sigma$ were rejected as outliers. Our best-fit orbital solution for C/1844 Y1, which we found to be in agreement with the published one by the MPC and JPL Horizons at the $1\sigma$ level, is presented in Table \ref{tab:orb}, together with our solutions for C/2019 Y4.

\begin{deluxetable*}{lc|cc|cc|cc}
\rotate
\tablecaption{Best-Fit Orbital Solutions for Comet Pair C/1844 Y1 and C/2019 Y4 (Heliocentric Ecliptic J2000.0)
\label{tab:orb}}
\tablewidth{0pt}
\tablehead{
\multicolumn{2}{c|}{}  & 
\multicolumn{2}{c|}{C/1844 Y1}  & 
\multicolumn{4}{c}{C/2019 Y4} \\  \cline{5-8}
\multicolumn{2}{c|}{Quantity}  & 
\multicolumn{2}{c|}{}  & 
\multicolumn{2}{c|}{Gravity-Only} &
\multicolumn{2}{c}{Nongravitational} \\  \cline{3-8}
 & &
Value & 1$\sigma$ Uncertainty & 
Value & 1$\sigma$ Uncertainty & 
Value & 1$\sigma$ Uncertainty
}
\startdata
Perihelion distance (au) & $q$
       & 0.250355 & 3.63$\times$10$^{-4}$ 
       & 0.252828387 & 9.66$\times$10$^{-7}$ 
       & 0.25281721 & 4.23$\times$10$^{-6}$\\ 
Eccentricity & $e$
       & 0.998910 & 5.78$\times$10$^{-4}$ 
       & 0.99924798 & 1.94$\times$10$^{-6}$
       & 0.99918998 & 1.28$\times$10$^{-6}$ \\ 
Inclination (\degr) & $i$
       & 45.5615 & 1.02$\times$10$^{-2}$ 
       & 45.382208 & 4.67$\times$10$^{-4}$
       & 45.386566 & 6.08$\times$10$^{-4}$ \\ 
Longitude of ascending node (\degr) & $\Omega$
                 & 120.6146 & 2.80$\times$10$^{-2}$ 
                 & 120.570633 & 4.36$\times$10$^{-4}$
                 & 120.574731 & 5.68$\times$10$^{-4}$ \\ 
Argument of perihelion (\degr) & $\omega$
                 & 177.4665 & 7.26$\times$10$^{-2}$ 
                 & 177.408982 & 3.73$\times$10$^{-4}$
                 & 177.411655 & 1.56$\times$10$^{-4}$ \\ 
Time of perihelion (TDB)\tablenotemark{$\dagger$} & $t_\mathrm{p}$
                  & 1844 Dec 14.18922 & 1.67$\times$10$^{-3}$ 
                  & 2020 May 31.012757 & 1.90$\times$10$^{-4}$
                  & 2020 May 31.020035 & 1.49$\times$10$^{-4}$ \\
Nongravitational parameters (au d$^{-2}$) & $A_{1}$
                  & N/A & N/A
                  & N/A & N/A
                  & $+$2.255$\times$10$^{-7}$ & 1.27$\times$10$^{-8}$ \\
 & $A_{2}$
                  & N/A & N/A
                  & N/A & N/A
                  & $-$3.05$\times$10$^{-8}$ & 1.01$\times$10$^{-8}$ \\ \hline
\multicolumn{2}{l|}{~~~Osculation epoch (TDB)} 
& \multicolumn{2}{c|}{JD 2395001.5 = 1845 Mar 11.0}
& \multicolumn{2}{c|}{JD 2458941.5 = 2020 Apr 02.0}
& \multicolumn{2}{c}{JD 2458941.5 = 2020 Apr 02.0 } \\
\multicolumn{2}{l|}{~~~Observed arc}
& \multicolumn{2}{c|}{1844 Dec 24-1845 Mar 11}
& \multicolumn{2}{c|}{2020 Jan 01-Mar 01}
& \multicolumn{2}{c}{2020 Jan 01-Apr 02} \\
\multicolumn{2}{l|}{~~~Number of observations\tablenotemark{$\ddagger$}}
& \multicolumn{2}{c|}{70 (10)}
& \multicolumn{2}{c|}{71 (33)}
& \multicolumn{2}{c}{104 (0)} \\
\multicolumn{2}{l|}{~~~Mean RMS residual (\arcsec)}
& \multicolumn{2}{c|}{$\pm15.521$}
& \multicolumn{2}{c|}{$\pm0.112$}
& \multicolumn{2}{c}{$\pm0.126$}
\enddata
\tablenotetext{\dagger}{The corresponding uncertainties are in days.}
\tablenotetext{\ddagger}{The unbracketed number is the number of observations used for the orbit determination, whereas the bracketed is the number of observations rejected as outliers. Note that, however, in the gravity-only solution for C/2019 Y4, the rejected observations are all from 2020 March 13 to April 02. See Section \ref{ssec_od} for details.}
\end{deluxetable*}

\subsection{Split Dynamics}
\label{ssec_sd}

\begin{deluxetable*}{lc|cc|cc}
\tablecaption{Orbital Elements for Comet Pair C/1844 Y1 and C/2019 Y4 (Solar System Barycentric Ecliptic J2000.0)
\label{tab:orb_bary}}
\tablewidth{0pt}
\tablehead{
\multicolumn{2}{c|}{Quantity}  & 
\multicolumn{2}{c|}{C/1844 Y1}  & 
\multicolumn{2}{c}{C/2019 Y4} \\  \cline{3-6}
 & &
Value & 1$\sigma$ Uncertainty & 
Value & 1$\sigma$ Uncertainty
}
\startdata
Periapsis distance (au) & $q$
       & 0.252465 & 3.58$\times$10$^{-4}$ 
       & 0.252501953 & 9.75$\times$10$^{-7}$ \\ 
Eccentricity & $e$
       & 0.998958 & 5.82$\times$10$^{-4}$ 
       & 0.99912756 & 1.93$\times$10$^{-6}$ \\ 
Inclination (\degr) & $i$
       & 45.3351 & 1.01$\times$10$^{-2}$ 
       & 45.339816 & 4.68$\times$10$^{-4}$ \\ 
Longitude of ascending node (\degr) & $\Omega$
                 & 120.5342 & 2.90$\times$10$^{-2}$ 
                 & 120.514602 & 4.37$\times$10$^{-4}$ \\ 
Argument of periapsis (\degr) & $\omega$
                 & 177.4367 & 7.30$\times$10$^{-2}$ 
                 & 177.450748 & 3.72$\times$10$^{-4}$ \\ 
Time of periapsis (TDB)\tablenotemark{$\dagger$} & $t_\mathrm{p}$
                  & 1844 Dec 14.4162 & 1.90$\times$10$^{-2}$ 
                  & 2020 May 31.903804 & 2.21$\times$10$^{-4}$
\enddata
\tablenotetext{\dagger}{The corresponding uncertainties are in days.}
\tablecomments{The orbital elements of the two comets are osculated from the gravity-only orbits in Table \ref{tab:orb} to a common osculation epoch of JD 2341972.5 $=$ TDB 1700 January 1.0. }
\end{deluxetable*}

The similarity between the orbits of C/1844 Y1 and C/2019 Y4 obviously hints at a possible genetic relationship between the two comets that they are likely two components that split from a common progenitor. Our primary goal is to investigate when the split event between the comet pair most likely took place and how large the separation speed was. 

We adopted a simplistic two-body dynamical model for the split event as follows. At some epoch $t_{\rm frg}$, the progenitor of the comet pair experienced a split event, during which two major components -- C/1844 Y1 and C/2019 Y4 were produced. The gravitational interaction between the pair was neglected. So was gravitational perturbation from the major planets in the solar system, as this effect is generally relatively unimportant \citep{2016ApJ...823....2S}. To make sure that this choice is valid to both C/1844 Y1 and C/2019 Y4, we created 1,000 clones for either of the comets based on the obtained best-fit gravity-only orbital elements and the covariance matrices. Then we utilised {\tt MERCURY6} \citep{1999MNRAS.304..793C} to integrate the clones together with the nominal orbits backward to 7 kyr ago, well past the previous perihelia that took place $\sim$5 kyr ago. What we found is that neither of the two comets had close approaches to any of the major planets since their last perihelion passages, validating the choice of neglecting planetary perturbation as an approximation. Our task is essentially equivalent to identifying how the orbits of C/1844 Y1 and C/2019 Y4 intersect and when the pair were both at the intersection point.

The major difference in the orbital elements of comets C/1844 Y1 and C/2019 Y4 lies in the epochs of their perihelion moments, whereas all other elements are not distinct (Table \ref{tab:orb}). Therefore for the simplistic dynamical model, we concentrated on the time interval of the perihelion epochs of the comet pair. As the orbit of C/2019 Y4 is more accurate, we used its elements, including periapsis distance $q$, eccentricity $e$, inclination $i$, longitude of ascending node $\Omega$, and argument of periapsis $\omega$ at epoch TDB 1700 January 01.0, referenced to the solar system barycentric ecliptic J2000.0, for the sake that the barycentric orbital elements of both comets remain largely constant between two consecutive periapsis returns. In this reference system, the differences between all of the orbital elements but the periapsis epoch are even within the corresponding $1\sigma$ orbital element uncertainties of C/1844 Y1 (Table \ref{tab:orb_bary}). We focus on how the separation velocity between the two comets and the split epoch should be can produce a difference in periapsis epochs of $\left|\Delta t_{\rm p}\right| \approx 175.5$ yr in the following.

\begin{figure*}
\epsscale{1.0}
\begin{center}
\plotone{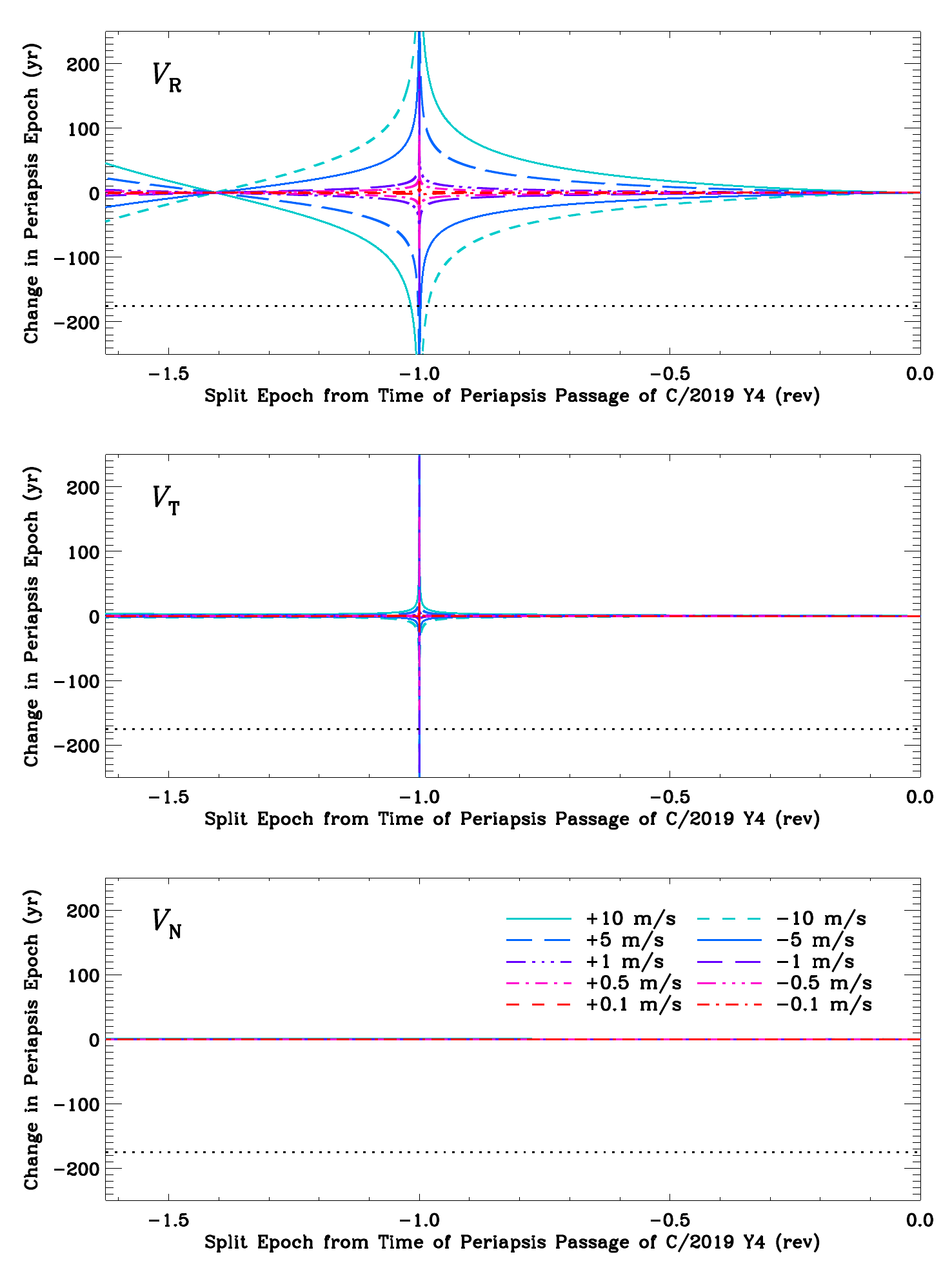}
\caption{
The change in periapsis epoch as a function of the split epoch with respect to the time of periapsis passage of C/2019 Y4 (expressed in numbers of orbital revolutions) and the separation velocity in terms of its radial (top), transverse (middle), and normal (bottom) components, given by our simplistic two-body dynamical model. In each panel, the actual difference between the periapsis epochs of the comet pair C/1844 Y1 and C/2019 Y4 is plotted as a horizontal black dotted line. As indicated in the legend, the results with different RTN separation velocity components are distinguished by colours and line styles. Note from the bottom panel that no out-of-plane components of the separation velocity $V_{\rm N}$ alone within the known range of the separation speeds \citep[$\sim$0.1-10 m s$^{-1}$;][and citations therein]{2004come.book..301B} can bring about significant changes in the periapsis epoch. We prefer that the split event most likely occurred around the previous perihelion return, which was $\sim$5 kyr ago, and that the gap between the periapsis passages of the pair is due to the in-plane component of the separation velocity.
\label{fig:YY_2bfrg}
} 
\end{center} 
\end{figure*}

Separation speeds between fragments of split comets are found to be in a range of $0.1 \la V_{\rm sep} \la 10$ m s$^{-1}$ \citep[][and citations therein]{2004come.book..301B}, of which only the in-plane component can alter the periapsis epoch, but not other orbital elements \citep{2016ApJ...823....2S}. We investigated the influence of the separation velocity in terms of radial, transverse, and normal (RTN) components upon the difference between the periapsis epochs of two fragments in the simplistic two-body dynamical model. The RTN coordinate system has its origin at the primary fragment, with the radial axis pointing away from the solar system barycentre, the normal axis directed along the total angular momentum of the primary fragment, and the transverse axis constructed to form a right-handed orthogonal system. At some split epoch $t_{\rm frg}$, the state vector of the primary fragment was computed from the orbital elements. The total velocity of the secondary fragment was updated by adding the separation velocity ${\bf V}_{\rm sep}$ to the velocity of the primary fragment. Thereby a new state vector was obtained, which was then converted into the orbital elements for the secondary fragment. 

We searched for conditions which should be satisfied for the split between C/1844 Y1 and C/2019 Y4 from the latest possible orbital revolution, such that the observed gap between the periapsis moments of the comet pair can be achieved. Given the periodicity of their orbits, there is an indefinite number of desired conditions from more than a single orbital revolution ago. The reason why we consider only the latest possible orbital revolution for the split event is as follows. Firstly, comet C/2019 Y4 is being fragmenting in the current apparition, indicative of the nucleus as a loosely bound aggregate as typical cometary nuclei \citep[e.g.,][and citations therein]{2004come.book..337W}. Secondly, given that the number of the observed split events of long-period comets was $\sim$30 amongst a total number of $\sim$10$^{3}$ long-period comets discovered in the past $\sim$150 yr (identified using the JPL Small-Body Database Search Engine), we can estimate a lower limit to the splitting rate as $\sim$2\% per century for each long-period comet, which is comparable to the one for short-period comets \citep[$\sim$3\% per century per comet;][]{2004come.book..301B}. Assuming that all long-period comets behave alike, within one orbit ($\sim$5 kyr) around the Sun, their progenitor (or any of its descendants) would experience at least one split event. Therefore, limiting our search for the split between comets C/1844 Y1 and C/2019 Y4 only within a timeframe not too much more than one orbital revolution in the past is a reasonable confinement. 

In Figure \ref{fig:YY_2bfrg}, we plot the change in the periapsis epoch $\Delta t_{\rm p}$ as a function of the RTN components of the separation velocity and the split epoch. We can learn that only if the split event that produced C/1844 Y1 and C/2019 Y4 occurred around the previous perihelion passage, which was $\sim$5 kyr ago, and the in-plane component of the separation velocity between the pair was $\ga$1 m s$^{-1}$, can the observed difference between the periapsis epochs of the pair be caused by separation speeds within the known range of split comets. Another conclusion we can draw is that the out-of-plane component of the separation velocity between the pair alone cannot bring about the observed periapsis epoch difference, which is similar to what \citet{2016ApJ...823....2S} found for another comet pair C/1988 F1 (Levy) and C/1988 J1 (Shoemaker-Holt).

To ensure that the planetary perturbation will not drastically alter the conclusion we drew based on the simplistic two-body model, we also employed our code, which includes planetary perturbation and has been utilised to analyse the split event of active asteroid P/2016 J1 (PANSTARRS) \citep{2017AJ....153..141H}, to find a best-fit nonlinear least-squared solution to the split parameters between C/1844 Y1 and C/2019 Y4. We treated C/2019 Y4 as the major component, and used its nominal orbit. The major difference here from the application in \citet{2017AJ....153..141H} is that, instead of fitting a list of topocentric astrometric observations, we fitted the heliocentric orbital elements of C/1844 Y1 (Table \ref{tab:orb}) with the associated covariance matrix from Section \ref{ssec_od}, and minimised the following quantity
\begin{equation}
\chi^{2} \left(t_{\rm frg}, V_{\rm R}, V_{\rm T}, V_{\rm N} \right) = \Delta {\bf E}^{\rm T} {\bf W} \Delta {\bf E}
\label{eq_chi2},
\end{equation}
\noindent where  $\Delta {\bf E}$ is the orbital element residual vector and ${\bf W}$ is the weight matrix determinable from the covariance matrix of the orbital elements ${\bf E} = \left(q, e, i, \Omega, \omega, t_{\rm p} \right)$ of C/1844 Y1 \citep[e.g.,][]{2010tod..book.....M}. Using different initial guesses, we soon realised that unless adopting an exhaustive and extensive search, which will be extremely time consuming, the code would converge to different solutions, indicative of the existence of multiple local minima. We present two of the best-fitted solutions we obtained are listed in Table \ref{tab:frg}. Although there is no definite solution to the split parameters between C/1844 Y1 and C/2019 Y4 from the N-body dynamical model, we think that the conclusion based on the simplistic two-body model remains valid with inclusion of planetary perturbation, because the best-fit solutions all have in-plane components of the separation velocity $\ga$1 m s$^{-1}$ and the split epochs around the previous perihelion of C/2019 Y4, at heliocentric distance $r_{\rm H} \la 10$ au.

\begin{deluxetable*}{lc|c|c}
\tablecaption{Fragmentation Solutions for Comet Pair C/1844 Y1 and C/2019 Y4
\label{tab:frg}}
\tablewidth{0pt}
\tablehead{
\multicolumn{2}{c|}{Quantity}  & 
Solution I  & Solution II
}
\startdata
Split epoch (TDB)\tablenotemark{$\dagger$} & $t_{\rm frg}$
       & B.C. 2901 Sep $2 \pm 20$ 
       & B.C. 2902 Mar $16 \pm 6$ \\ 
RTN separation velocity (m s$^{-1}$) & & \\
~~~~Radial component & $V_{\rm R}$
       &  $-3.1 \pm 0.1$ 
       & $+1.9 \pm 0.1$ \\ 
~~~~Transverse component & $V_{\rm T}$
       & $+5.0 \pm 0.3$ 
       & $-0.2 \pm 0.3$ \\ 
~~~~Normal component & $V_{\rm N}$ 
       & $0 \pm 0$
       & $+0.1 \pm 1.9$ \\  \hline
Goodness of fit & $\chi^{2}$
                 & 11.3 
                 & 8.3 \\
O-C residuals\tablenotemark{$\ddagger$} & & \\
~~~~Perihelion distance (au) & $\Delta q$
                  & $-5.38 \times 10^{-4}$
                  & $-2.13 \times 10^{-5}$\\
~~~~Eccentricity & $\Delta e$
                  & $-1.45 \times 10^{-4}$ 
                  & $-1.47 \times 10^{-4}$ \\
~~~~Inclination (\degr) & $\Delta i$
                  & $-4.20 \times 10^{-3}$ 
                  & $-6.14 \times 10^{-3}$ \\
~~~~Longitude of ascending node (\degr) & $\Delta \Omega$
                  & $+1.90 \times 10^{-2}$ 
                  & $+1.84 \times 10^{-2}$ \\
~~~~Argument of perihelion (\degr) & $\Delta \omega$
                  & $-3.99 \times 10^{-2}$ 
                  & $-1.67 \times 10^{-2}$ \\
~~~~Time of periapsis (d) & $\Delta t_\mathrm{p}$
                  & $-2.08 \times 10^{-4}$ 
                  & $-1.39 \times 10^{-3}$ \\
\enddata
\tablenotetext{\dagger}{The corresponding uncertainties are in days.}
\tablenotetext{\ddagger}{The differences between the observed and computed orbital elements of C/1844 Y1.}
\tablecomments{Only the nominal orbit of C/2019 Y4 was used for the computation, and therefore the uncertainties of the split parameters presented in the table must have been seriously underestimated. We have fixed $V_{\rm N} = 0$ in Solution I, of which the split epoch would place the separation between C/1844 Y1 and C/2019 Y4 at a heliocentric distance of $r_{\rm H} \approx 4$ au postperihelion. The split epoch of Solution II corresponds to the split event at $r_{\rm H} \approx 4$ au preperihelion. Compared to the orbital period ($\sim$5 kyr), we think that the conclusion from the simplistic two-body dynamic model that the split event occurred around the previous perihelion return and the magnitude of the in-plane component of the separation velocity $V_{\rm sep} \ga 1$ m s$^{-1}$ is validated.}
\end{deluxetable*}

Although the fragmentation mechanism that led to the disruption between the pair remains unclear, we can safely rule out the possibility of tidal disruption, because the perihelion distance is larger than the Roche radius of the Sun for comets (a few solar radii) by at least an order of magnitude, and there was no close approach to any of the major planets in the timeframe we investigated either. We postulate that possible fragmentation mechanisms include rotational instability due to anisotropic mass loss, excessive internal thermal stress or gas pressure \citep[and citations therein]{2004come.book..301B}, and overwhelming differential stress due to dynamic sublimation pressure \citep{2015Icar..258..430S}. A detailed discussion of the potential physical mechanisms that caused the observed ongoing disintegration event of C/2019 Y4 will be presented in another paper in preparation.


\section{Summary}
\label{sec_sum}

The key conclusions of our study on long-period comet C/2019 Y4 (ATLAS), the sibling of C/1844 Y1 (Great Comet) are listed as follows:

\begin{enumerate}

\item C/2019 Y4 was observed to brighten intrinsically with a growth rate in the effective scattering cross-section $\left(2.0 \pm 0.1 \right) \times 10^{2}$ m$^{2}$ s$^{-1}$ from January 2020, until $\sim$70 d prior to its perihelion passage, whereafter the comet started to fade in brightness and started to lose its central condensation.

\item The colour of the comet across the $g$ and $r$ bands once turned even bluer ($g - r \approx 0.2$) than that of the Sun in late March 2020 from an earlier colour that was slightly red ($g - r \approx 0.6$) in January and February 2020 before the fade in brightness occurred. This is likely due to that a massive amount of previously buried fresh volatiles suddenly exposed to sunlight.

\item With the tailward-bias corrected astrometric observations, we detected an enormous radial nongravitational effect in the heliocentric motion of the comet, $A_{1} = \left(+2.25 \pm 0.13 \right) \times 10^{-7}$ au d$^{-2}$. The transverse nongravitational parameter, $A_{2} = \left(-3.1 \pm 1.0\right) \times 10^{-8}$ au d$^{-2}$, is far less significant. Altogether, we conclude that C/2019 Y4 has disintegrated since mid-March 2020.

\item The split between the comet pair C/1844 Y1 and C/2019 Y4 occurred around the previous perihelion passage of the progenitor, with the magnitude of the in-plane component of their separation velocity $\ga$1 m s$^{-1}$.

\item We estimate that the nucleus of C/2019 Y4 was $\ga$60 m in radius before the observed disintegration event. The dominant grains in the dust tail in March 2020 had $0.03 \la \beta \la 0.1$ (corresponding mean dust radii $\sim$10-40 \micron, assuming a bulk density of $\rho_{\rm d} = 0.5$ g cm$^{-3}$) and were ejected protractedly, with ejection speed $\sim$30 m s$^{-1}$ in early of the month, and increased to $\sim$80 m s$^{-1}$ at the end for grains of $\beta \sim 0.1$, similar to those of other long-period comets.

\end{enumerate}

\acknowledgements
{
We thank Jon Giorgini, Bill Gray, and Paul Wiegert for helpful discussions, Xing Gao for obtaining the NEXT images, Ana Hayslip, Casey Kyte, Ishara Nisley, and LaLaina Shumar for assistance in acquiring the LDT data, David Tholen for sharing us with details of his astrometric measurements, and the anonymous reviewer for their insightful comments. The operation of Xingming Observatory was made possible by the generous support from the Xinjiang Astronomical Observatory of the Chinese Academy of Sciences. NEXT is funded by the Ningbo Bureau of Education and supported by the Xinjiang Astronomical Observatory. LDT is operated at Lowell Observatory, a private, nonprofit institution dedicated to astrophysical research and public appreciation of astronomy. Lowell operates the LDT in partnership with Boston University, the University of Maryland, the University of Toledo, Northern Arizona University and Yale University. The Large Monolithic Imager was built by Lowell Observatory using funds provided by the National Science Foundation (AST-1005313). The LDT observations were obtained by the University of Maryland observing team, consisted of L.~M. Feaga, Q.-Z. Ye, J.~M. Bauer, T.~L. Farnham, C.~E. Holt, M.~S.~P. Kelley, J.~M. Sunshine, and M.~M. Knight.

}

\vspace{5mm}
\facilities{4.3 m LDT, 0.6 m NEXT}

\software{{\tt AstroImageJ} \citep{2017AJ....153...77C}, {\tt FindOrb}, IDL, {\tt MERCURY6} \citep{1999MNRAS.304..793C}, {\tt PHOTOMETRYPIPELINE} \citep{2017A&C....18...47M}}




\end{document}